\documentclass[12pt]{iopart}
\usepackage[latin1]{inputenc}
\usepackage[T1]{fontenc}
\usepackage[english]{babel}
\usepackage{graphicx}
\usepackage{latexsym}
\begin{document}
\title[Dynamical study of the hyperextended scalar-tensor theory...]{Dynamical study of the hyperextended scalar-tensor theory in the empty Bianchi type I model.}
\author{Stéphane Fay}
\address{66 route de la Montée Jaune\\
37510 Savonnières\\FRANCE\footnote{Steph.Fay@Wanadoo.fr}}
\begin{abstract}
The dynamics of the hyperextended scalar-tensor theory in the empty Bianchi type I model is investigated. We describe a method giving the sign of the first and second derivatives of the metric functions whatever the coupling function. Hence, we can predict if a theory gives birth to expanding, contracting, bouncing or inflationary cosmology. The dynamics of a string inspired theory without antisymetric field strength is analysed. Some exact solutions are found.
\\
\\
\\
Published in Classical and Quantum Gravity copyright 2000 IOP Publishing Ltd\\
Classical and Quantum Gravity, Vol 17, 7, 2000.\\
http://www.iop.org
\end{abstract}
\maketitle
\section{Introduction} \label{s1}
We study the dynamics of the metric functions for the hyperextended scalar-tensor theory in the empty Bianchi type I model.

The cosmological principle is based on the hypothesis of an isotropic and homogeneous Universe. However, at early times, Universe could have been anisotropic. We can quote several reasons in favour of this hypothesis \cite{a}. Firstly, the isotropic hypothesis rests on observations such as the cosmological background. But it does not rule out the possibility of an anisotropic Universe for primordial time. Secondly, if the Universe is too isotropic and homogeneous, it is difficult to explain formation of structures, like galaxies: presence of anisotropies is necessary. Last, it could be easier to avoid singular Universe under these conditions.

Anisotropic Universes are described by the Bianchi models. Among these models, the only ones which isotropize and are in accordance with our present Universe at late time, are these of type $I$, $V$, $VII_0$, $VII_h$ and $IX$. Current observations favour open and flat models and recent measurements seem to indicate that our present Universe undergoes inflation \cite{9}. Then it is a serious possibility that our Universe be spatially flat. It corresponds to the Bianchi type I model which will be the geometrical framework of this paper. 

An important field of study in cosmology is the introduction of scalar fields in gravity theories. There are many reasons to justify their presence. Firstly, they are predicted by unified theories and could be the result of the compactification of extra dimensions appearing in theories like supersymetric, Kaluza-Klein or string theories. Secondly, they provide a way to get inflation \cite{6}, ending naturally without any fine-tuning. At last, the scalar-tensor theories can respect the solar system tests \cite{11} as well as nucleosynthesis one but make very different predictions from General Relativity at early time.

Among the scalar tensor theories, the most famous and simplest generalisation of General Relativity is the Brans-Dicke theory \cite{b}. The coupling between the graviton and the dilaton, represented by the scalar field $\phi$, is described by a coupling constant $\omega$. If it is larger than 500, the theory respects the solar system tests. However, string theory in the low energy limit, which could describe the physics of the early Universe, is identical to Brans-Dicke theory with $\omega=-1$ after scalar field redefinition. Such contradiction between these two values of the coupling constant looks like the cosmological constant ($\Lambda$) problem: its observed value is about 120 orders smaller than what expected from a theoretical point of view. One way to solve this problem is to choose a variable cosmological constant. We can adopt the same solution concerning the coupling constant and consider a coupling function depending on the scalar field, $\omega(\phi)$. Such theories are called Generalised Scalar-Tensor theories (GST) and have been studied in the FLRW \cite{c} and anisotropic \cite{d} models in presence of matter.

In these theories $\phi^{-1}$ plays the role of a varying gravitational constant. However such a choice seems to be arbitrary. It is interesting to consider a function $G(\phi)^{-1}$ instead of $\phi$ in front of the scalar curvature term in the Lagrangian: this is the Hyperextended Scalar-Tensor theory (HST) \cite{1}\cite{2}. It can be rewritten as a GST \cite{f} by redefining a scalar field $\Phi=G(\phi)^{-1}$. Then we need to find the inverse function of $G(\phi)^{-1}$ which is not always analytically defined. This justifies the study of the HST.

Let's write few words about the relations between GST and HST and their relationship with General Relativity. The GST are agreed with the solar system tests if at late time $\omega\rightarrow \infty$ and $\omega_{\phi}\omega^{-3}\rightarrow 0$. For the HST, there is an additional unknown function of the scalar field, $G(\phi)^{-1}$. If we put $\Phi=G(\phi)^{-1}$, we obtain a GST with a coupling function written $\Omega(\phi)$. It can be expressed as a function of $\omega(\phi)$ and $G(\phi)^{-1}$: $\Omega(\Phi)=\omega(\phi)G (\phi)^{-1}(G^{-1}_\phi)^{-2}\phi^{-1}$. Then, we deduce that the two conditions so that HST is agreed with solar system tests will be respectively: $\omega G^{-1}(G^{-1}_\phi)^{-2}\phi^{-1}\rightarrow \infty$ and $(G^{-1}_\phi)^3 G^{2}\omega^{-2}\phi^2(\omega_\phi\omega^{-1}+G^{-1}_\phi G-\phi^{-1}-2G^{-1}_{\phi\phi}G)\rightarrow 0$. If we choose $G(\phi)^{-1}=\phi$, we recover the usual conditions for GST.

Lots of gravitation theories belong to HST class as dilaton gravity with $G^{-1}=1/2e^{-\phi}$ and $\omega=-1/2\phi e^{-\phi}$, generally coupled scalar field with $G^{-1}=1/2(\gamma-\xi\phi^2)$ and $\omega=1/2\phi$, induced gravity with $G^{-1}=1/2\epsilon\phi^2$ and $\omega=1/2\phi$, etc \cite{e}. It is difficult to choose physically interesting $G^{-1}$ and $\omega$. Different periods of the Universe could be approximated by different coupling functions. A way to select them is to impose that the theory be in accordance with the solar system tests at late time. We can also use dynamical criterions: the metric functions should be increasing at late time, eventually have a minimum so that they avoid the Big Bang, and have an inflationary period.

It is in view of determining such characteristics for the metric functions that we will examine the dynamics of the HST in the empty. A more realistic model will take into account matter fields. But then, most of time only asymptotic studies are workable for a given form of $\omega$ and $G^{-1}$. Generally it does not allow to detect the presence of several extrema, quasi-static phases for the dynamics or multiple inflationary phases. Our motivation is also to detect such physically important behaviours for any form of $\omega$ and $G^{-1}$, i.e. to study the dynamics of the metric functions whatever the value of the time and not only asymptotically. The price to pay for this full description of the dynamics is the absence of matter fields.

However, since their presence tends to oppose to expanding Universe, we hope that necessary and sufficient conditions we will establish to get expansion, inflation or quasi-static phases for instance in an empty model, will be either necessary or sufficient if matter fields are present. Hence, more complete studies of large classes of new theories specified by $\omega$ and $G^{-1}$ with matter fields could be stimulated if they already have physically interesting dynamical characteristics in the empty. At the opposite, large classes of theories could be discriminated if their dynamical behaviours in the empty were in contradiction with current observations.

The paper is organised as follows: in section \ref{s2}, we write the field equations in the empty Bianchi type I model and introduce new variables to transform them into a differential system of first order. We give the exact solution of the field equations. In section \ref{s3}, we study the sign of the first derivatives of the metric functions and determine in which conditions they are increasing, decreasing or have extrema. In section \ref{s4}, we study their second derivatives to predict the appearance of inflation or quasi-static phases. In these two last sections, we applied our results to a string inspired theory without H-field. We conclude in section \ref{s5} by showing the advantages of the method we present in this work to study any empty HST. We give the conditions on $G(\phi)^{-1}$ and $\omega(\phi)$ so that the Universe respects the solar system tests, be in expansion and accelerated at late time, and avoid the Big-Bang.
\section{Field equations and exact solution} \label{s2}
\subsection{Field equations} \label{ss21}
We use the following line element:
\begin{equation} \label{0}
ds^{2}=-dt^{2}+e^{2\alpha} (\omega^1)^2+e^{2\beta}(\omega^2)^2+e^{2\gamma}(\omega^3)^2
\end{equation}
where the $\omega^{i}$:
\begin{eqnarray}
\omega^1=dx\nonumber\\
\omega^2=dy\nonumber\\
\omega^3=dz\nonumber\\\nonumber
\end{eqnarray}
are the 1-forms of the Bianchi type I model, $t$ the proper time and $e^\alpha$, $e^\beta$, $e^\gamma$ the metric functions depending on $t$. The Lagrangian of the HST is written:
\begin{equation} \label{1}
L=G(\phi)^{-1}R-\frac{\omega(\phi)}{\phi}\phi_{,\mu}\phi^{,\mu}
\end{equation}
$G$ and $\omega$ depend on the scalar field and specify the theory. Varying the action with respect to the space-time metric and scalar field, we obtain the field equations and the Klein-Gordon equation:
\begin{equation} \label{2}
R_{\mu\nu}-\frac{1}{2}g_{\mu\nu}R=G\left[\frac{\omega}{\phi}\phi_{,\mu}\phi_{,\nu}-\frac{\omega}{2\phi}\phi_{,\lambda}\phi^{\lambda}g_{\mu\nu}+(G^{-1})_{,\mu;\nu}-g_{\mu\nu}\Box (G^{-1})\right]
\end{equation}
\begin{equation} \label{3}
\dot{\phi}^{2}\left[-\frac{\omega_{\phi}}{\phi}+\frac{\omega}{\phi^{2}}-G(G^{-1})_{\phi}\frac{\omega}{\phi}\right]+\frac{2\omega}{\phi}\Box \phi+3G(G^{-1})_{\phi}\Box (G^{-1})=0
\end{equation}
a dot meaning a derivative with respect to t time. Using the form (\ref{0}) of the metric and $\tau$ time defined by $dt=e^{\alpha+\beta+\gamma}d\tau$, we get:
\begin{eqnarray} \label{4}
 &\alpha''+\alpha' G(G^{-1})' +\frac{1}{2}G(G^{-1})''=0& \label{4a} \\
 &\beta''+\beta' G(G^{-1})' +\frac{1}{2}G(G^{-1})''=0& \\
 &\gamma''+\gamma' G(G^{-1})' +\frac{1}{2}G(G^{-1})''=0& \\
 &\alpha'\beta'+\alpha'\gamma'+\beta'\gamma'+G(G^{-1})'(\alpha'+\beta'+\gamma')-\omega\frac{G}{2}\frac{\phi^{,2}}{\phi}=0& \\ \nonumber
\end{eqnarray}
\begin{equation} \label{5}
\phi^{,2}\left[-\frac{\omega_{\phi}}{\phi}+\frac{\omega}{\phi^{2}}-G(G^{-1})_{\phi}\frac{\omega}{\phi}\right]-2\omega\frac{\phi''}{\phi}-3G(G^{-1})_{\phi}(G^{-1})''=0
\end{equation}
a prime meaning a derivative with respect to $\tau$. The functions $\alpha$, $\beta$ and $\gamma$ play equivalent roles in the field equations. So, in what follows, we will only consider the metric function $e^\alpha$.

We are interested in the signs of first and second derivatives of the metric functions and not in their amplitudes. Since the product $e^{\alpha+\beta+\gamma}$ is positive, the signs of the first derivatives in the $\tau$ and $t$ times will be the same whereas they will be different for second derivatives. Hence, to determine the sign of $(e^{\alpha})^.$, we will study this of $\alpha'$ in section \ref{s3}. In section \ref{s4} we will determine the signs of the second derivatives by studying separately $(e^{\alpha })^{..}$ and $(e^{\alpha})''$. This is justified by the fact that sometimes solutions are known in the $\tau$ time and not in the $t$ time.

Now, we define new variables $A$, $B$, $C$ and $F$ in order to transform the second order field equations into a first order system:
\begin{eqnarray}
 &A=\alpha' G^{-1} & \nonumber\\
 &B=\beta' G^{-1} & \label{6}\\
 &C=\gamma' G^{-1} & \nonumber\\ 
 &F=\frac{1}{2}(G^{-1})' & \nonumber\\ \nonumber
\end{eqnarray}
Then, after integration, the spatial components of the field equations are written:
\begin{eqnarray} \label{7}
& A+F=A_0& \label{7a}\\
& B+F=B_0&\\
& C+F=C_0&\label{7c}\\ \nonumber
\end{eqnarray}
$A_0$, $B_0$ and $C_0$ being integration constants. We also integrate the Klein-Gordon equation and get:
\begin{equation} \label{8}
\frac{3}{4}(G^{-1})^{,2}+\frac{1}{2}G^{-1}\omega\phi^{-1}\phi^{,2}=-\Pi
\end{equation}
$\Pi$ being an integration constant. This last relation is written again:
\begin{equation} \label{9}
\left[\frac{3}{4}(G^{-1})_\phi^{\mbox{ }2}+\frac{G^{-1}\omega}{2\phi}\right]\phi^{,2}=-\Pi
\end{equation}
From the constraint equation of the field equations we deduce the following relation between the integration constants:
\begin{equation} \label{10}
A_0B_0+A_0C_0+B_0C_0=-\Pi
\end{equation}
The quantity between square brackets in the left hand-side of equation (\ref{9}) is proportional and has the same sign as the energy density of the scalar field in the Einstein frame. For physical reasons, we will take a positive energy density, i.e.
\begin{equation} \label{13}
\frac{3}{4}(G^{-1})_\phi^{\mbox{ }2}+\frac{G^{-1}\omega}{2\phi}>0
\end{equation}
Hence, we deduce that $-\Pi>0$. If we choose $G^{-1}=\phi$, we recover the usual relation for a positive energy density for GST, i.e. $3+2\omega>0$. The sign of $\phi'$ is constant and depends on the sign of the square root of the energy density: if we take it positive (negative), the scalar field will be increasing (decreasing). Hence, the scalar field being a monotone function of time, it will be considered as a time variable.

From now, we will just consider the first spatial component of the field equations, i.e. equation (\ref{7a}) since we only need to study the dynamics of $e^{\alpha}$. The set of values $(A,F)$, solution of (\ref{7a}), can be graphically represented in the $(A,F)$ plane by a straight line. During time evolution, the dynamics of the solution is described by the motion of a point of coordinate $(A,F)$ on this set.
\subsection{Exact solution} \label{ss22}
Using (\ref{7a}) and the first relation of (\ref{6}), we deduce the exact solution for $\alpha(\tau)$:
\begin{equation} \label{11}
\alpha-\alpha_0=\int \frac{A_0}{G^{-1}}d\tau-\frac{1}{2}ln(G^{-1})
\end{equation}
$\alpha_0$ being an integration constant. If we write $d\tau=\phi^{,-1}d\phi$ and express $\phi'$ using (\ref{9}), we obtain $\alpha(\phi)$:
\begin{equation} \label{12}
\alpha-\alpha_{0}=\int \frac{A_0}{G^{-1}}\sqrt{-\frac{1}{\Pi}\left[\frac{3}{4}(G^{-1})_\phi ^{\mbox{ }2} +\frac{G^{-1}\omega}{2\phi}\right]}d\phi-\frac{1}{2}ln(G^{-1})
\end{equation}
and analogous relations for $\beta$ and $\gamma$ with couples of constants $(\beta_0,B_0)$ and $(\gamma_0,C_0)$ respectively instead of $(\alpha_0,A_0)$.\\
\\
There are two interesting asymptotical values for the couple $(A,F)$. The first one is $(A,F)\rightarrow (0,A_0)$. It means that $G^{-1}\rightarrow 2(A_0\tau+A_1)$. Then, we deduce from (\ref{11}) that the metric function tends toward a constant. Thus, the point $(0,A_0)$ stands for the static solution for $e^\alpha$. The second one is $(A,F)\rightarrow (A_0,0)$. Then $G^{-1}$ tends toward a constant. From (\ref{11}) we get that $\alpha\rightarrow \alpha_1\tau+\alpha_2$, $\alpha_1$ and $\alpha_2$ being some constants. The function $\beta$ and $\gamma$ will behave in the same way in respectively $(B_0,0)$ and $(C_0,0)$. In the t time, this solution for the metric functions corresponds to power laws of t.
\section{First derivatives of the metric functions} \label{s3}
Using (\ref{12}), we can write $\alpha'$ as a function of $\phi$ and then study its sign. However, even in the case of very simple functions $G^{-1}$ and $\omega$, the expression thus obtained is often difficult to analyse. The method we describe below allow to get in a simple manner the sign of the first derivative.
\subsection{Sign of the first derivative} \label{ss31}
\begin{figure}[h]
\begin{center}
\includegraphics{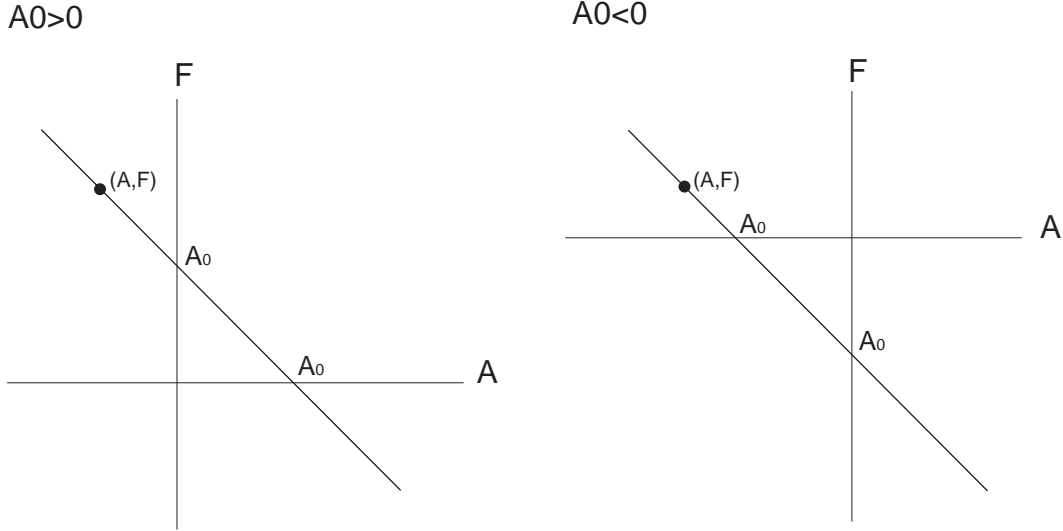}
\end{center}
\caption{The straight line describing the set of solutions $(A,F)$ of the first spatial component of the field equations in the plane $(A,F)$. To know the sign of the first derivative of the metric function $e^{\alpha}$, we have to analysed the dynamics of a point $(A,F)$ on this straight line. For that, we split the scalar field variation interval, considered as a time variable, in n sub-intervals such as the signs of $G^{-1}$, $(G^{-1})'$ and $(G^{-1})''$ be constant. Hence, on each of this interval $\left[\phi_{l-1},\phi_l\right]$, we know the sign of $F$ given by $(G^{-1})'$ and in which direction the point $(A,F)$ moves on the straight line depending on the fact that $F$ increases or decreases, i.e. on the sign of $(G^{-1})''$. Then we have to check if $F$ can take the value $2A_0$ when $\phi\in\left[\phi_{l-1},\phi_l\right]$. When it is true, it means that a metric function have an extremum for this range of the scalar field. Otherwise, it is monotone. Thus we deduce what is the sign of $A$ on this scalar field interval and, as $G^{-1}$ has also a constant sign, we get the sign of $\alpha'=AG$. Hence, on each interval $\left[\phi_{l-1},\phi_l\right]$, we obtain the sign of the first derivative of the metric function.}
\label{fig1}
\end{figure}

Now, we are explaining how to determine the sign of the first derivative of $\alpha$ for successive intervals of scalar field, considered as a time variable. For the clarity of the discussion, we will assume that $\phi$ is an increasing function of $t$ or $\tau$ time, i.e. $\sqrt{-\Pi}>0$. Moreover, we will need to evaluate $(G^{-1})'$ and $(G^{-1})''$ for some values of the scalar field. To this end, we express the derivatives of $G^{-1}$ with respect to $\tau$ as functions of $\phi$. Since $(G^{-1})'=G^{-1}_\phi\phi'$, we obtain:
\begin{equation} \label{15}
(G^{-1})'=(G^{-1})_\phi \sqrt{\frac{-\Pi}{\frac{3}{4}(G^{-1}_{\mbox{   }\phi})^2+\frac{G^{-1}\omega}{2\phi}}}
\end{equation}
In the same way, we get:
\begin{equation} \label{14}
(G^{-1})''=-4\Pi\frac{2(G^{-1})_{\phi\phi}\omega G^{-1}\phi-(G^{-1})_\phi^{\mbox{   }2}\omega\phi+(G^{-1})_\phi(\omega G^{-1}-G^{-1}\omega_\phi \phi)}{(2G^{-1}\omega+3\phi(G^{-1})_\phi^{\mbox{  }2})^2}
\end{equation}
To apply our method we need also to determine the following intervals:
\begin{enumerate}
\item The scalar field variation interval is defined by the condition (\ref{13}): its energy density in the Einstein frame have to be positive. We write it as $\left[\phi_0,\phi_n\right]$.
\item We split it in several subintervals such as in each of them, $G^{-1}$, $(G^{-1})'$ and $(G^{-1})''$ have constant signs. We note these subintervals $\left[\phi_0,\phi_n\right]= \left[\phi_0,\phi_1\right]\cup... \left[\phi_{l-1},\phi_l\right] \cup ...\cup \left[\phi_{n-1},\phi_n\right]$. 
\end{enumerate}
Remember that they can be compared to time intervals since $\phi$ is an increasing function of time.\\
\textbf{As a first step}, we have to determine the direction of the motion of the point $(A,F)$ on the straight line defined by equation (\ref{7a}) (see figure \ref{fig1}). Since $F=1/2(G^{-1})'$, it means that on each interval $\left[\phi_{l-1},\phi_l\right]$ when $(G^{-1})''>0$, $F$ increases and thus the point $(A,F)$ moves from the right to the left. Otherwise, $F$ decreases and the points moves from the left to the right.\\
\textbf{As a second step}, we determine the sign of $A$ on each interval $\left[\phi_{l-1},\phi_l\right]$. Lets illustrate this point when $A_0>0$:
\begin{itemize}
\item If $(G^{-1})'<0$, $F$ is negative. We see on the straight line represented on figure \ref{fig1} that then $A>0$ whatever the sign of $(G^{-1})''$.
\item If $(G^{-1})'>0$ and $(G^{-1})''>0$, $F$ is positive and increases on $\left[\phi_{l-1},\phi_l\right]$: $F\in\left[1/2(G^{-1})'(\phi_{l-1}),1/2(G^{-1})'(\phi_{l})\right]$. Since the sign of $A$ changes when $F=A_0$, we have to check if this value belongs or not to this last interval. We have three possibilities:
	\begin{itemize}
\item If $(G^{-1})'(\phi_{l})<2A_0$, it implies that $(G^{-1})'(\phi_{l-1})<2A_0$ and then $A>0$.
\item If $(G^{-1})'(\phi_{l-1})>2A_0$, it implies that $(G^{-1})'(\phi_{l})>2A_0$ and then $A<0$.
\item If $(G^{-1})'(\phi_{l-1})<2A_0$ and $(G^{-1})'(\phi_{l})>2A_0$, as $F$ increases, first we have $A>0$ and then $A<0$.
	\end{itemize}
\item If $(G^{-1})'>0$ and $(G^{-1})''<0$, $F$ is positive and decreases. Then, for the same reasons as before, we have three possibilities:
	\begin{itemize}
\item If $(G^{-1})'(\phi_{l-1})<2A_0$, it implies that $(G^{-1})'(\phi_{l})<2A_0$ and then $A>0$.
\item If $(G^{-1})'(\phi_{l})>2A_0$, it implies that $(G^{-1})'(\phi_{l-1})>2A_0$ and then $A<0$.
\item If $(G^{-1})'(\phi_{l-1})>2A_0$ and $(G^{-1})'(\phi_{l})<2A_0$, as $F$ decreases, first we have $A<0$ and then $A>0$.
	\end{itemize}
\end{itemize}
Hence, this shows that the sign of $A$ when the point $(A,F)$ moves on the straight line of figure \ref{fig1} representing the solution of the equation (\ref{7a}), is perfectly determined on each interval $\left[\phi_{l-1},\phi_l\right]$ by the sign of $(G^{-1})'$, $(G^{-1})''$ and the ordering of the quantities $(G^{-1})'(\phi_{l-1})$, $(G^{-1})'(\phi_{l})$ and $2A_0$. Of course, the method is the same if the scalar field is decreasing or $A_0<0$.\\
\textbf{As a third and last step}, we determine the sign of $\alpha'$ on each intervals $\left[\phi_{l-1},\phi_l\right]$. Since the signs of A and $G^{-1}$ are known on $\left[\phi_{l-1},\phi_l\right]$, we immediately deduce the sign of $\alpha'=AG$: If $G>0$ ($G<0$), when $A>0$, the metric function is increasing (decreasing). Otherwise, it is decreasing (increasing).\\
\\
Thus, on each interval $\left[\phi_{l-1},\phi_l\right]$ we are able to determine if the metric function is increasing, decreasing or have an extremum. The scalar field being a monotone function of time, we can describe for any time the evolution of the sign of the first derivative of $e^\alpha$, i.e. its dynamics.\\
\\
What happens when the theory respects the solar system tests? Our present Universe being in expansion, we write the conditions so that a metric function is increasing depending on $G^{-1}$, $\omega$ and their derivatives with respect to $\phi$. Since $A=A_0-F=\alpha'G^{-1}$, the metric function is increasing on an interval of scalar field if $(A_0-1/2(G^{-1})')G>0$. Moreover, we know that $\Omega=\omega G^{-1}(G^{-1}_{\phi})^{-2}\phi^{-1}$ have to diverge at late time so that the theory is compatible with the relativistic values of the PPN parameters. If we examine the relation (\ref{15}), we deduce that this limit corresponds to $(G^{-1})'\rightarrow 0$. Hence, for a theory respecting the solar system tests at late time, the metric function $\alpha$ will be increasing if $A_0G>0$. Since gravitation is attractive and $G$ can play the role of an effective gravitational constant, we have $G^{-1}>0$ and thus $A_0>0$. Finally, as $(G^{-1})'=2F\rightarrow 0$, we deduce also that all the metric functions tend toward power law in the t time as shown previously. 

To summarise, for an expanding Universe, respecting the solar system tests at late time, all the metric functions tends toward power laws of the proper time and the initial conditions are such as $(A_0,B_0,C_0)>(0,0,0)$. Mathematically, it would be interesting to transform the system of equations (\ref{7a}-\ref{7c}) so that we use the dynamical system methods and analyse if power laws solutions for the metric functions correspond to future attractor. Such a study is beyond the scope of this paper and will be done in a next one.

In what follows, it seems to be physically reasonable to impose $G^{-1}>0$. For the GST, it is equivalent to choose $\phi>0$.
\subsection{Applications} \label{ss32}
\subsubsection{Brans-Dicke theory} \label{sss311}
\begin{figure}[h]
\begin{center}
\includegraphics{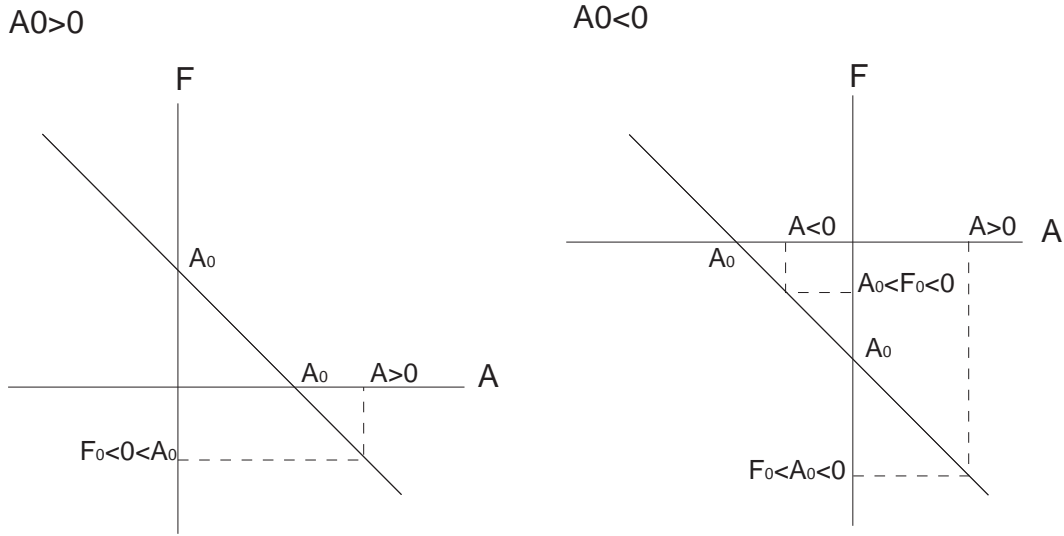}
\end{center}
\caption{The Brans-Dicke theory. We have sign of $(G^{-1},(G^{-1})',(G^{-1})'')=(+,-,0)$ for any value of $\phi$. F is a negative constant equal to $F_0$. Whatever the sign of $A_0$, when $F<A_0$, $A>0$. If $A_0<0$, when $F>A_0$, then $A<0$. The sign of $A$ is the same as the first derivative of $e^\alpha$.}
\label{fig2}
\end{figure}

We chose:
\begin{eqnarray} \label{16}
G^{-1}=e^{-\phi}\nonumber\\
\omega=\omega_0 \phi e^{-\phi}\nonumber
\end{eqnarray}
with $\omega_0>-3/2$ so that the energy density of the scalar field is positive. This choice corresponds to the string theory without H-field and with the term $\omega_0$ in front of $\phi_{,\mu}\phi^{,\mu}$ instead of 1. By putting $\Phi=e^{-\phi}$, we recover the Lagrangian of the Brans-Dicke theory with a coupling constant equal to $-\omega_0$. We calculate that:
\begin{eqnarray} \label{17}
 &(G^{-1})''=0& \nonumber\\
 &(G^{-1})'=-2\sqrt{-\Pi/(3+2\omega_0)}& \nonumber
\end{eqnarray}
The sign of $\alpha'$ is the same as $A$ since $G^{-1}>0$. $F$ is a negative constant equal to $F_0=-\sqrt{-\Pi/(3+2\omega_0)}$. If $F_0>A_0$ then $(A,F)$ is such as $A<0$ and the metric function is decreasing. It is increasing otherwise (figure \ref{fig2}). We recover the usual dynamics of the Brans-Dicke theory for the Bianchi type $I$ model.
\subsubsection{String inspired theory} \label{sss322}
We modify the previous Lagrangian. In string theory, we can take into account the string loop effects by substituting the coupling function $e^{-\phi}$ by the series  $B(\phi)=e^{-\phi}+a_0+a_1 e^{\phi}+a_2 e^{2\phi}+...$ In our application, we will limit the series to its two first terms \cite{4}. Note that no theory predicts the value of the $a_i$ today. Hence, cosmological applications are susceptible to restrict the range of these parameters. Moreover, we consider again that $\omega_0$ can be different from $1$. This is justified by the fact that in our four dimensional Universe, it could exist moduli fields whose forms depend on the compactification scheme, producing $\omega_0\not=1$ \cite{3}. Recent progress have been made on dual transformations concerning empty string theory (i.e. without any axion or moduli fields) with a constant $\omega_0$ \cite{5}.\\
We examine the string inspired theory without H-field defined by:
\begin{eqnarray} \label{18}
G^{-1}=e^{-\phi}+n\nonumber\\
\omega=\omega_0\phi(e^{-\phi}+n)
\end{eqnarray}
Using (\ref{12}), we have calculated the exact solution of the field equations (see \ref{a1}). It is clearly easier to use the method described above than derive the sign of $\alpha'$ from this solution. As $\phi$ is increasing, the theory will respect the solar system tests at late time for $\phi\rightarrow +\infty$. Then $G^{-1}\rightarrow n$ and $n$ can be seen as the present value of the gravitational constant.

We search for the scalar field variation interval so that $G^{-1}$ and its energy density is positive. This last quantity vanishes for $e^{\phi_{1,2}}=1/n(-1\pm\sqrt{-3/(2\omega_0)})$. After few algebra we obtain the table \ref{tab1} giving all the possible scalar field variation intervals depending on $n$ and $\omega_0$.

We find that the sign of $(G^{-1})'$ is always negative and conclude that $F<0$ whatever $\phi$. The sign of $(G^{-1})''$ is the same as $n\omega_0$: it means that $2F=(G^{-1})'$ is a monotone function. Hence, the signs of $(G^{-1},(G^{-1})',(G^{-1})'')$ are these of $(+,-,n\omega_0)$: they are constant whatever $\phi$ and we have no need to split the scalar field variation interval in n sub-intervals.

This last result and the "step 2", show that we have to compare the values of $(G^{-1})'$ to the constant $2A_0$ when $\phi$ is equal to the boundaries of each of its variation interval so that we detect the presence or absence of extrema. We calculate that:\\
$(G^{-1})'(+\infty,-\infty,\phi_{1,2},-ln(-n))=(0,-\sqrt{-4\Pi(3+2\omega_0)^{-1}},-\infty,-\sqrt{-4\Pi/3})$.

\begin{figure}[h]
\begin{center}
\includegraphics{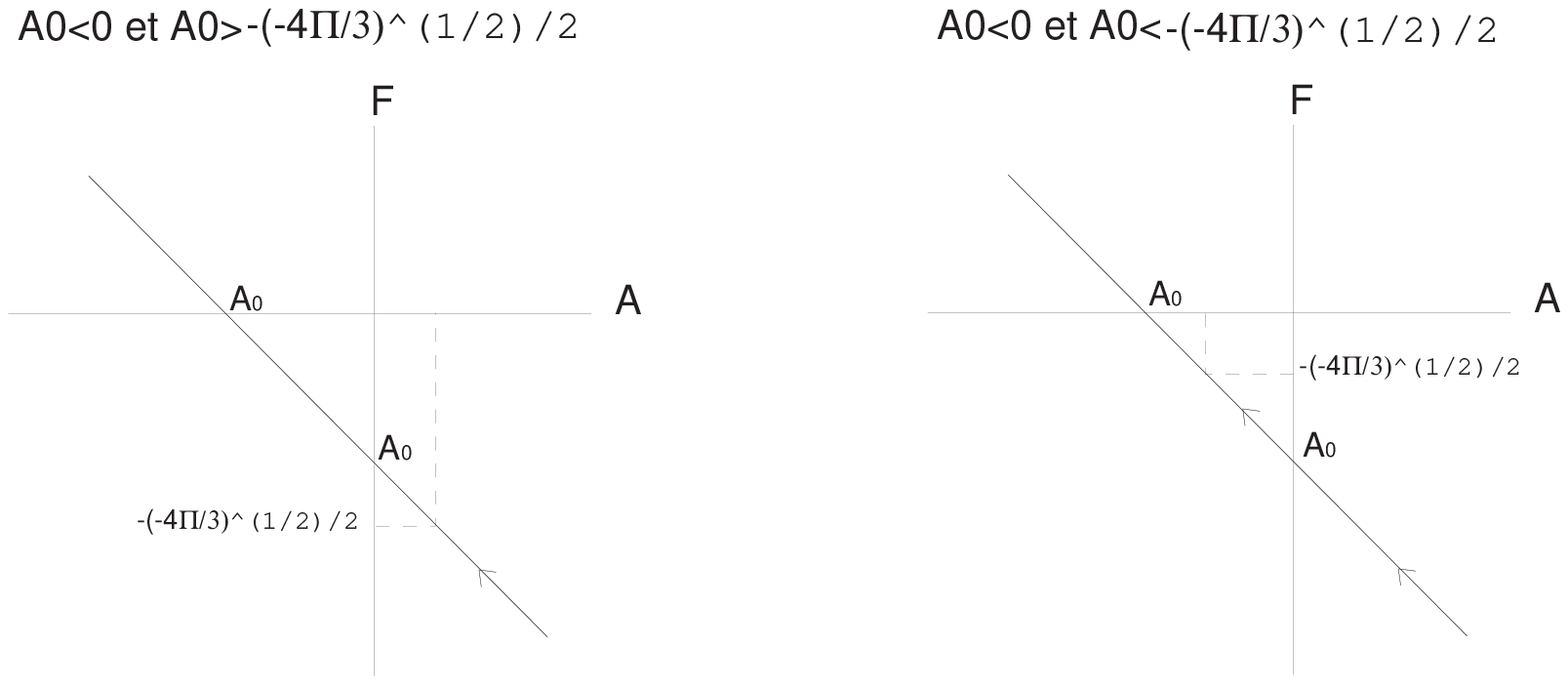}
\end{center}
\caption{The string inspired theory when $\omega<-3/2$, $n<0$ and $A_0<0$. $F$ increases (this is indicates by the direction of the arrows on each straight line of the figures) from $-\infty$ to $-\sqrt{-4\Pi/3}/2$ since $(G^{-1})''>0$.  On the first figure, this last value is smaller than $A_0$. Then $A$ is always positive and since $G^{-1}>0$, $\alpha'$ is positive. The metric function $e^\alpha$ is increasing. On the second figure, $-\sqrt{-4\Pi/3}/2$ is larger than $A_0$. As long as $F<A_0$, $A>0$ and then when $F>A_0$, $A<0$. Since $A$ and $\alpha'$ have the same sign we deduce that the metric function has a maximum.}
\label{fig3}
\end{figure}

From these results, we use the method described in section \ref{s3} to get the dynamics of the metric function $e^\alpha$:
\begin{itemize}
\item If $A_0>0$:
	\begin{itemize}
	\item As $2F=(G^{-1})'<0$, $A$ is always positive. Since $G^{-1}>0$, it follows that the metric function 	$e^{\alpha}$ is always increasing.
	\end{itemize}
\item If $A_0<0$:
	\begin{itemize}
	\item If $\omega_0<-3/2$ and $n<0$, $2F=(G^{-1})'$ increases from $-\infty$ to $-\sqrt{-4\Pi/3}$. If this last 	value is smaller than $2A_0$, $e^\alpha$ is increasing. Otherwise it has a maximum. This case is shown on 	figure \ref{fig3}.
	\item If $\omega_0\in\left[-3/2,0\right]$ and $n>0$, $(G^{-1})'$ decreases from $-\sqrt{-4\Pi/(3+2\omega_0)}$ to 	$-\infty$. If the first value is smaller than $2A_0$, $e^\alpha$ is increasing. Otherwise, it has a minimum.
	\item If $\omega_0\in\left[-3/2,0\right]$ and $n<0$, $2F=(G^{-1})'$ increases from $-\sqrt{-4\Pi/(3+2\omega_0)}$ to 	$-\sqrt{-4\Pi/3}$. If the two values are smaller than $2A_0$, $e^\alpha$ is increasing. If both are larger than 	$2A_0$, $e^\alpha$ is decreasing. If $2A_0$ belongs to the interval defined by these values, $e^\alpha$ 	has a maximum.
	\item If $\omega_0>0$ and $n>0$, $2F=(G^{-1})'$ increases from $-\sqrt{-4\Pi/(3+2\omega_0)}$ to $0$. If the first 	value is larger than $2A_0$, $e^\alpha$ is decreasing, otherwise a maximum exists. It is the only case for 	which at late time, the	metric function tends toward a power law type for $t$. Moreover, the theory is 	compatible with solar system tests since $\phi\rightarrow +\infty$. However, the dynamics at late time is 	not in accordance with the observations. On this simple example, we see that conditions for the respect of 	the solar system tests are not sufficient to ensure a realistic dynamics of the Universe for our present time.
	\item If $\omega_0>0$ and $n<0$, $2F=(G^{-1})'$ decreases from $-\sqrt{-4\Pi/(3+2\omega_0)}$ to $-\sqrt{-	4\Pi/3}$. If these two values are smaller than $2A_0$, $e^\alpha$ is increasing. If they are larger than 	$2A_0$, $e^\alpha$ is decreasing. If $2A_0$ belongs to the interval defined by these values, it has a minimum.
	\end{itemize}
\end{itemize}
Hence the method described in the previous section allows to know all the conditions for which the metric functions are decreasing, increasing or "bouncing". It would have been more difficult to get the same results from the exact solution $\alpha(\phi)$.
\section{Sign of the second derivative} \label{s4}
In this section, we study the sign of the second derivatives of $e^\alpha$ in $\tau$ and $t$ times. This is justified by the fact that sometimes solutions are known in one time but not in the other. In what follows, we assume that we know the scalar field variation interval.
\subsection{Sign of the second derivative in the $\tau$ time} \label{ss41}
The sign of the second derivative of the metric function in the $\tau$ time is the same as:
\begin{equation} \label{18a}
G^{-2}(e^\alpha)''=G^{-2}e^\alpha(\alpha''+\alpha^{,2})
\end{equation}
The spatial component (\ref{4a}) of the field equations provides:
\begin{equation} \label{19}
G^{-2}\alpha''=-A(G^{-1})'-1/2G^{-1}(G^{-1})''
\end{equation}
From the equation (\ref{7a}) we get:
\begin{equation} \label{19a}
G^{-2}\alpha^{,2} = A^2=(A_0-1/2(G^{-1})')^2
\end{equation}
Then, from the two last equations we deduce that the sign of $(e^\alpha)''$ is the same as:
\begin{equation} \label{20a}
G^{-2}(\alpha''+\alpha^{,2})=\frac{3}{4}(G^{-1})^{,2}-2A_0(G^{-1})'-\frac{1}{2}G^{-1}(G^{-1})''+A_0^2
\end{equation}
It is a second-degree equation for $(G^{-1})'$. With the help of the relations (\ref{15}-\ref{14}), we can express its coefficients as a function of the scalar field. Its determinant is equal to:
\begin{eqnarray} \label{21}
\Delta &=& A_0^2  + 6 \Pi G^{-1} \mbox{[}-G^{-1} \omega (G^{-1})_\phi + \phi \omega (G^{-1})_\phi^{\mbox{  }2}  + \phi G^{-1} (G^{-1})_{\phi} \omega_\phi -\nonumber\\
&&2 \phi G^{-1} \omega (G^{-1})_{\phi\phi}\mbox{]}/(2 G^{-1} \omega + 3 \phi (G^{-1})_\phi^{\mbox{  }2} )^2\\ \nonumber
\end{eqnarray}
and its roots are:
\begin{equation} \label{22}
(G^{-1})'_{root^1_2}=\frac{4A_0\pm\sqrt{\Delta}}{3}
\end{equation}
We deduce that:
\begin{itemize}
\item If $\Delta<0$, the second-degree equation is negative and then $(e^{\alpha})''>0$.
\item If $\Delta>0$, $(G^{-1})'-(G^{-1})'_{root1}$ and $(G^{-1})'-(G^{-1})'_{root2}$ have different signs, $(e^{\alpha})''<0$.
\item If $\Delta>0$, $(G^{-1})'-(G^{-1})'_{root1}$ and $(G^{-1})'-(G^{-1})'_{root2}$ have the same signs, $(e^{\alpha})''>0$.
\end{itemize}
All these inequalities can be expressed as some functions of $\phi$. From them, it is possible to derive the scalar field intervals so that $(e^{\alpha})''$ is positive or negative. Since we can determine the scalar field intervals for which the sign of $(e^\alpha)'$ is constant, it is possible to describe completely the dynamical evolution of the metric function $\alpha(\tau)$.
\subsection{Sign of the second derivative in the $t$ time} \label{ss42}
The sign of the second derivative in the $t$ time is the same as:
\begin{equation} \label{23}
\frac{d^{2} e^{\alpha}}{dt^{2}}=\left[(e^{\alpha})''-(e^{\alpha})'(\alpha'+\beta'+\gamma')\right]e^{-2(\alpha+\beta+\gamma)}
\end{equation}
Using (\ref{18a}) to express $G^{-2}(e^{\alpha})''$ and the relations (\ref{15}-\ref{14}), we get the expression giving the sign of $(e^{\alpha})^{..}$:
\begin{eqnarray} \label{24}
G^{-2}\frac{d^2 e^\alpha}{dt^2}e^{\alpha+2(\beta+\gamma)} &=& (Bo + Co) \left[-Ao + \frac{(G^{-1})_{\phi}}{2}\sqrt{-\frac{\Pi}{\frac{G^{-1}\omega}{2\phi}+\frac{3(G^{-1})_\phi^{\mbox{   }2}}{4}}}\right] \nonumber \\
&&-2\Pi G^{-1} \mbox{[}-G^{-1} \omega (G^{-1})_{\phi} + \phi \omega (G^{-1})_{\phi}^{\mbox{  }2}  + 
 \phi G^{-1} (G^{-1})_{\phi} \omega_\phi \nonumber \\
&&- 2 \phi G^{-1} \omega (G^{-1})_{\phi\phi}\mbox{]}
 / (2 G^{-1} \omega + 3 \phi (G^{-1})_{\phi}^{\mbox{  }2} )^2\\ \nonumber
\end{eqnarray}
Since it can be written as a function of the scalar field, it is possible to deduce the scalar field intervals so that $(e^{\alpha})^{..}$ is positive or negative. By Comparing them with these for which the sign of the first derivative is constant, we will get the qualitative dynamical behaviour of $\alpha(t)$. When the sign of the second derivative of the metric function with respect to $t$ is positive on a scalar field interval, the dynamics is accelerated. If, at the same time\footnote{Here, we consider $\phi$ as the time variable.}, the metric function is increasing, we are in the presence of inflation. Lets note that it happens naturally without any potential. Such phenomenon has been studied in the GST and received the name of kinetic inflation \cite{6}. Inflation in the HST has been studied in \cite{7}. When the right hand side of the equation (\ref{24}) vanishes, the metric function $e^{\alpha}$ have a point of inflection. Physically, it means that we could be in presence of a quasi-static phase for the dynamics of the Universe, at least in the direction associated with the metric function $e^\alpha$.

If we assume that the theory is in agreement with solar system tests at late time, then we know that $\Omega\rightarrow \infty$ and $(G^{-1})'\rightarrow 0$. We introduce this limit in the expression (\ref{24}) and obtain the condition to have an inflationary behaviour for the metric function at late time: $2\Pi G^{-1} \mbox{[}-G^{-1} \omega (G^{-1})_{\phi} + \phi \omega (G^{-1})_{\phi}^{\mbox{  }2}  +  \phi G^{-1} (G^{-1})_{\phi} \omega_\phi - 2 \phi G^{-1} \omega (G^{-1})_{\phi\phi}\mbox{]} / (2 G^{-1} \omega + 3 \phi (G^{-1})_{\phi}^{\mbox{  }2} )^2<-A_0(B_0+C_0)$. If our present Universe undergoes inflation (observations of higher redshift objects seem to be necessary to confirm this phenomenon \cite{10}), this last inequality could play the same role as the two conditions necessary so that a GST respects the solar system test (i.e. $\omega>500$ and $\omega_\phi\omega^{-3}\rightarrow 0$) and thus, help to select physical interesting HST.
\subsection{Application in the $t$ time: the string inspired theory} \label{ss43}
For this theory, (\ref{24}) takes the form:
\begin{eqnarray} \label{25}
4\Pi n\omega_0 e^{\phi}(1+ne^{\phi})^2(2\omega_0 n^2 e^{2\phi}+4\omega_0 n e^{\phi}+3+2\omega_0)^{-2}+\nonumber \\
(B_0+C_0)\left[-A_0-\sqrt{-2\Pi(2\omega_0n^2 e^{2\phi}+4\omega_0 n e^{\phi}+3+2\omega_0)^{-1}}\right]
\end{eqnarray}
We look for the asymptotic sign of this equation for the different cases described in table \ref{tab1} and depending on $n$ and $\omega_0$. From this table, we deduce:
\begin{itemize}
\item If $\omega_0<-3/2$ and $n<0$, $e^\alpha$ is decelerated at early time. At late time the second derivative has the sign of $-(B_0+C_0)(A_0+\sqrt{-\Pi/3})$.
\item If $\omega_0>-3/2$ and $n<0$, the second derivative at early and late times has respectively the sign of $-(B_0+C_0)(A_0+\sqrt{-\Pi/(2\omega_0+3)})$ and $-(B_0+C_0)(A_0+\sqrt{-\Pi/3})$.
\item If $\omega_0\in\left[-3/2,0\right]$ and $n>0$, the second derivative at early time has the sign of $-(B_0+C_0)(A_0+\sqrt{-\Pi/(2\omega_0+3)})$. The dynamics of the metric function is accelerated at late time. Since we have shown that it is always increasing for these values of $\omega_0$ and $n$, we have inflation.
\item If $\omega_0>0$ and $n>0$, the second derivative at early time has the sign of $-(B_0+C_0)(A_0+\sqrt{-\Pi/(2\omega_0+3)})$ whereas at late time it has this of $-A_0(B_0+C_0)$.
\end{itemize}
Lets note that if $A_0>0$, $e^\alpha$ is always an increasing function and any accelerated behaviour will correspond to inflation. Moreover, (\ref{25}) can be seen as a polynome for $e^\phi$. We do not make its complete study since this section is just an application but it seems to be clear that it should have more than one zero. Hence, the theory should have several phases of inflation. Mathematically, an asymptotical study could not have detected such behaviour. This is one of the advantage of the method presented in this paper.
\section{Conclusion} \label{s5}
We have studied the dynamical behaviour of the metric functions for the HST in the empty Bianchi type I model for any form of $G^{-1}$ and $\omega$. Such dynamical study has always been done for the GST with matter field in Bianchi models \cite{c} and FLRW models \cite{d}. However, most of time it concerns asymptotic behaviours. Here, we have made the choice to consider a simpler theory, i.e. without matter field, but to study its dynamics for any time and not only asymptotically. Mathematically, we get a more accurate description of the dynamics than with asymptotical methods. The spliting of the scalar field variation interval in several ones allow to get all the extrema of the metric functions as well as their types. The calculation of the zeros of equation (\ref{24}) enable to get the intervals of $\phi$, considering as a time variable, in which a metric function is accelerated, decelerated as well as its inflexion points. Comparing these two types of scalar field intervals, we are able to describe completely the dynamical behaviour of the metric functions. Thus physically, it is possible to predict if a theory, defined by $\omega(\phi)$ and $G(\phi)$, will give birth to an Universe with several bounces. Such a scenario could be one of the keys to homogenise the Universe in the manner of a Mixmaster model. We can also predict if there will have several periods of inflation. It is also an interesting behaviour since some problems need inflation to be solved (age problem, isotropisation) whereas for others, it is prefered that the Universe be decelerated (formation structures). Last, we can detect quasi-static phases which are likely to favour the appearance of some structures we observe in the Universes and to solve the age problem. We think that the detection of such characteristics in an empty model may stimulate and justify more complex researches when matter fields are present. Asymptotical studies are generally not able to detect such behaviours.

We have applied this method to a string inspired theory. Since we have determined the exact solution of the field equations as function of $\phi$ (cf \ref{12}), it is easy to calculate the exact solution of this theory (\ref{a1}). Clearly, it seems to be difficult to study the dynamical behaviour of the metric function from the expression thus obtained. We have shown that the late time behaviour of this theory is not compatible with our observed Universe. However it could be an interesting model for early time behaviour. The metric functions are monotone or have one and only one extremum. If all the metric functions have a minimum, the Big-Bang can be avoided. For that, it is necessary that $\omega_0\in\left[-3/2,0\right]$ and $n>0$ or $\omega_0>0$ and $n<0$. We have also shown that several periods of inflation are possible.

Although GST is often claimed to be equivalent to HST it is only true if the inverse function of $G^{-1}$ can be analytically determined \cite{2}. Hence, HST is a more general class of scalar tensor theories than GST which is obtained for $G^{-1}=\phi$. In this case, it is well known that the theory will respect the solar system tests if $\omega\rightarrow \infty$ and $\omega_\phi\omega^{-3}\rightarrow 0$. For any form of $G$, these conditions become $\omega G^{-1}(G^{-1}_\phi)^{-2}\phi^{-1}\rightarrow \infty$ and $(G^{-1}_\phi)^3 G^{2}\omega^{-2}\phi^2(\omega_\phi\omega^{-1}+G^{-1}_\phi G-\phi^{-1}-2G^{-1}_{\phi\phi}G)\rightarrow 0$. For this limit, we have shown that the metric functions tend toward a power law type in the $t$ time and $G^{-1}$ toward a constant which then may correspond to the present value of the gravitational constant. Mathematically, it would be interesting to study the fields equation (\ref{7a}-\ref{7c}), which are first order equations, in the light of the dynamical system methods \cite{14} so that we learn if this behaviour could be a late time attractor. Physically, the fact that $G$ tends toward a constant is associated with a power law types for the metric functions is in good agreement with what should be the dynamical behaviour of our present Universe and thus confirmed the viability of scalar tensor theories.

We conclude by giving the conditions on $G$ and $\omega$ so that the Universe at late time, respects the solar system tests, be accelerated and bouncing. When the theory respects the solar system tests, we have $(G^{-1})'\rightarrow 0$. Then, the Universe is in expansion and accelerated if $A_0G$ is positive, $2\Pi G^{-1} \mbox{[}-G^{-1} \omega (G^{-1})_{\phi} + \phi \omega (G^{-1})_{\phi}^{\mbox{  }2}  +  \phi G^{-1} (G^{-1})_{\phi} \omega_\phi - 2 \phi G^{-1} \omega (G^{-1})_{\phi\phi}\mbox{]} / (2 G^{-1} \omega + 3 \phi (G^{-1})_{\phi}^{\mbox{  }2} )^2<-A_0(B_0+C_0)$ and other similar conditions obtained by circular permutations on $A_0$, $B_0$ and $C_0$. Since $G$ could play the role of an effective gravitational constant, it means that today $G>0$ and thus $(A_0,B_0,C_0)>0$. This restricts the range of the initial conditions. If moreover we want that all the metric functions have a minimum so that the Big-Bang is avoided and if we assume that $G$ was positive (negative) at early time, we need to choose $G(\phi)^{-1}$ and $\omega(\phi)$ so that $(G^{-1})''$ is negative (positive). The conditions concerning the respect of the solar system tests and these described in this paragraph and concerning the dynamics of the metric functions put strong constraints on the form of $G(\phi)^{-1}$ and $\omega(\phi)$. For the latter, to our knowledge, we have not seen equivalent ones in the literature.

\appendix
\section{Exact solution of the string inspired theory} \label{a1}
From (\ref{12}), we get $\alpha(\phi)$:\\
\\
\underline{Solution with $\omega_0>0$}:
\begin{equation}
\alpha-\alpha_0= 
\end{equation}
$-ln(\sqrt{e^{-\phi+n}})+ Ao \{\sqrt{2} \sqrt{\omega_0} \phi + \sqrt{3} ln(e^{-\phi}+ n) - \sqrt{3 + 2 \omega_0} ln\{-\Pi(3 + 2 \omega_0 + 2 e^{\phi}    n \omega_0 + e^{\phi}    \sqrt{3 + 2 \omega_0}  \mbox{[}(3 + 2 \omega_0 + 4 e^{\phi}    n \omega_0 + 2 e^{2\phi} n^2  \omega_0) / (e^{2\phi} -\Pi)\mbox{]}^{1/2} \sqrt{-\Pi})/ e^{\phi}\} - \sqrt{3} ln\{-3-\Pi e^{-\phi} + \sqrt{3} \mbox{[}(3 + 2 \omega_0 + 4 e^{\phi}    n \omega_0 + 2 e^{2\phi}      n^2  \omega_0) /            (e^{2\phi} -\Pi)\mbox{]}^{1/2} (-\Pi)^{3/2}   \} +  \sqrt{2} \sqrt{\omega_0} ln\{-\omega_0 \Pi e^{-\phi} + -n \omega_0 \Pi +  -\Pi\sqrt{\omega_0/2} \mbox{[}(3 + 2 \omega_0 + 4 e^{\phi}    n \omega_0 +  2 e^{2\phi}      n^2  \omega_0) / (e^{2\phi}      -\Pi^{3/2})\mbox{]}^{1/2}\}\}/ (2 \sqrt{-\Pi})$\\
\\
with $\tau-\tau_0=\sqrt{-\Pi}(6 \phi + 2 \sqrt{6}\sqrt{\omega_0} arctan\{ (3 + 2 \omega_0 + 2 e^{\phi}    n \omega_0)/(\sqrt{6} e^{\phi}    n \sqrt{\omega_0})\} + 
  3 ln\{(3 + 2 \omega_0 + 4 e^{\phi}    n \omega_0 + 2 e^{2\phi}   n^2  \omega_0)e^{-2\phi}\}) / (3 n^2  \omega_0)$.\\
\\
\underline{Solution with $\omega_0<0$}:
\begin{equation}
\alpha-\alpha_0= 
\end{equation}
$-ln(\sqrt{e^{-\phi}+n})+ Ao \{\sqrt{-\omega_0} arctan\{(\sqrt{2} e^{\phi}    (1 + e^{\phi}    n) \sqrt{-\omega_0} 
   \mbox{[}(3 - 2 -\omega_0 - 4 e^{\phi}    n -\omega_0 - 2 e^{2\phi} n^2  -\omega_0)/(-\Pi e^{2\phi})\mbox{]}^{1/2}
\sqrt{-\Pi}) / (-3 + 2 -\omega_0 + 4 e^{\phi}    n -\omega_0 + 2 e^{2\phi} n^2  -\omega_0)\} / (\sqrt{-2\Pi}) + \sqrt{3}ln(e^{-\phi}+n)/(2\sqrt{-\Pi}) + ((-3 + 2 -\omega_0) ln\{-\Pi(3 - 2 -\omega_0 - 2 e^{\phi}    n -\omega_0 + 
            e^{\phi}    \sqrt{3 - 2 -\omega_0}\mbox{[}(3 - 2 -\omega_0 - 4 e^{\phi}    n -\omega_0 - 2 e^{2\phi} n^2  -\omega_0)/(-\Pi e^{2\phi})\mbox{]}^{1/2}\\
 \sqrt{-\Pi})e^{-\phi} \}) / (2 \mbox{[}-\Pi(3 - 2 -\omega_0)\mbox{]}^{1/2}) -  \sqrt{3} ln\{ -3-\Pi e^{-\phi} + \sqrt{3}
\mbox{[}(3 - 2 -\omega_0 - 4 e^{\phi}    n -\omega_0 - 2 e^{2\phi} n^2  -\omega_0)/(-\Pi e^{2\phi})\mbox{]}^{1/2} -\Pi^{3/2}\} / (2 \sqrt{-\Pi})\}$.\\
\\
with $\tau-\tau_0=\sqrt{-\Pi}(-6 \phi + 2 \sqrt{6} \sqrt{-\omega_0} 
 arctanh\{ (-3 - 2 \omega_0 - 2 e^{\phi}    n \omega_0)/(\sqrt{6} e^{\phi} n \sqrt{-\omega_0})\} -
3 ln\{(-3 - 2 \omega_0 - 4 e^{\phi}    n \omega_0 - 2 e^{2\phi}   n^2  \omega_0)e^{-2\phi}\}) / (-3 n^2  \omega_0)$

\newpage
\begin{table} 
\caption{Scalar field variation intervals such as its energy density and $G^{-1}$ be positives} 
\label{tab1}
\begin{center}
\begin{tabular}{llll} 
 & $\omega_0<-3/2$ & $\omega_0\in\left[-3/2,0\right]$ & $\omega_0>0$ \\
\hline
$n<0$ & $\phi\in\left[\phi_2,ln(-1/n)\right]$ & $\phi\in\left]-\infty,ln(-1/n) \right]$& $\phi\in\left]-\infty,ln(-1/n) \right]$\\
$n>0$ & $\mbox{energy density}
<0$ & $\phi\in\left]-\infty,\phi_1 \right]$ & $\phi\in\left]-\infty,+\infty\right[$\\
\end{tabular}
\end{center}
\end{table}

\section*{References}

\begin{thebibliography}{9}

\bibitem{a}
N. A. Batakis, A. A. Kehagias,
"Anisotropic space-times in homogeneous string cosmology",
Nucl.Phys. B{\bf449}, 248-264
(1995).

\bibitem{13}
V. Sahni, A. Starobinsky, 
"The case for a positive cosmological constant $\Lambda$-term", 
to appear in International Journal of Modern Physics, 
astro-ph/9904398, 
(1999).

\bibitem{6}
Janna Levin,
"Kinetic inflation in stringy and other cosmologies",
Phys Rev D{\bf51},
pp 1536,
(1995);
"Gravity-Driven acceleration of the cosmic expansion",
Phys Rev D{\bf51},
 462,
(1995).

\bibitem{11}
C. M. Will,
Phys. Reports {\bf113}, 345,
(1984).

\bibitem{b}
C. Brans and RH. Dicke,
Phys Rev {\bf124}, 925,
(1961).

\bibitem{c}
J.D.Barrow, P.Parson,
"The behaviour of cosmological models with varying-G",
Phys. Rev D{\bf55},
1906-1936,
(1997);
D. I. Santiago, D. Kalligas, R. V. Wagoner
"Scalar-Tensor cosmologies and their late time evolution",
Phys. Rev. D{\bf58},
(1998);
A. Navarro, A. Serna, J. M. Alimi,
"Asymptotic and exact solutions of perfect-fluid scalar tensor cosmologies",
Phys. Rev. D,
(1999);
A. Serna, J. M. Alini,
"Scalar-tensor cosmological models",
Phys. Rev. D{\bf53}, 3074-3086,
(1996).

\bibitem{d}
J.P. Mimoso, D. Wands,
"Anisotropic scalar-tensor cosmologies",
Phys Review D{\bf52}, p5612-5627,
(1995);
A. Billyard, A. Coley, J. Ib$\acute{a}\tilde{n}$ez,
"On the asymptotic behaviour of cosmological models in scalar-tensor theories of gravity",
Phys. Rev. D{\bf59}, 023507,
(1999).

\bibitem{1}
D. F. Torres, H. Vucetich,
"Hyperextended Scalar Tensor Gravity",
Phys. Rev. D{\bf54}, 7373-7377
(1996).

\bibitem{2}
D. F. Torres,
"Classes of Anisotropic Cosmologies of Scalar-Tensor Gravitation",
gr-qc/9612048.

\bibitem{f}
John D Barrow,
"Non singular scalar-tensor cosmologies",
Phys Review D{\bf48}, number 8,
(1993);
A. R. Liddle, D. Wands,
Phys. Rev. D{\bf45}, 2665,
(1992).

\bibitem{e}
Jai-Chen Hwang,
"Cosmological perturbations in generalised gravity theories: conformal transformation,
Class. Quantum Grav. {\bf14}, 1981-1991,
(1997).

\bibitem{3}
Sudipta Mukherji,
"A note on Brans-Dicke cosmology with axion",
Mod. Phys. Lett. A{\bf12}, 639-645,
(1997).

\bibitem{4}
M. C. Bento, O. Bertolami,
"Cosmological solutions of higher-curvature string effective theories with dilaton",
gr-qc/9503057;
T. Damour, A. M. Polyakov,
"String theory and gravity",
Gen. Rel. Grav. {\bf26}, 1171-1176,
(1994).

\bibitem{5}
J. E. Lidsey,
"On the cosmology and symmetry of dilaton-axion gravity",
gr-qc/9609063.

\bibitem{7}
Diego F. Torres,
"Slow Roll Inflation in Non-Minimally Coupled Theories: Hyperextended Gravity Approach",
Phys.Lett. A{\bf225}, 13-17,
(1997).

\bibitem{9}
S. Perlmutter et al, 
"Measurements of  $\Omega$ and $\Lambda$ from 42 Hight-Redschift Supernovae", 
Astrophysical Journal, 
(1999);
Riess et al, 
"Observational evidence from supernovae for an accelerating Universe and a cosmological constant", 
A. J. , {\bf 116}, 1009, 
(1998).

\bibitem{10}
G. Starkman, M. Trodden, T. Vachaspati,
"Observation of cosmic acceleration and determining the fate of the Universe",
astro-ph/9901405.

\bibitem{14}
"Dynamical systems in cosmology",
Edited by "J. Wainwright and G. F. R. Ellis",
Cambridge University Press, (1997).

\end{thebibliography}
\end{document}